\documentclass[letter, scriptaddress, twocolumn,  prb, showpacs,showkeys]{revtex4-1}

	\usepackage{amsmath}
           \usepackage{mathrsfs}
	\usepackage{amssymb}
	\usepackage{graphicx} 
	\usepackage{makeidx}
	\usepackage{amsfonts}
	\usepackage[ansinew]{inputenc}
	\usepackage[usenames, dvipsnames]{pstricks}
	\usepackage{subfigure}
	\usepackage{epsfig}
	\usepackage{pst-grad} 
	\usepackage{pst-plot} 
	\usepackage[colorlinks, hyperindex]{hyperref}
	\hypersetup
	{
		colorlinks, %
		citecolor=blue, %
		linkcolor=Green, %
		urlcolor=blue, %
	}

\makeindex

\begin{document}

\title{{\Large\bf Zeno-Gravity Correspondence\\Zeno's Dichotomy Paradox and Gravitational red-shift near Event Horizon}}

\author{Sridip Pal} 
\email{sridippaliiser@gmail.com; srpal@ucsd.edu}
\affiliation{Department of Physics\\
University of California, San Diego\\
9500 Gilman Drive, La Jolla, CA 92093, USA}
\author{Benjamin Grinstein}
\email{bgrinstein@ucsd.edu}
\affiliation{Department of Physics\\
University of California, San Diego\\
9500 Gilman Drive, La Jolla, CA 92093, USA}

\begin{abstract}
We relate Zeno's Dichotomy paradox with gravitational red-shift near event horizon in a spherically symmetric space-time. A dictionary of this connection, henceforth called as Zeno-Gravity correspondence, has been built up. The infinite sequence of Zeno's paradox gets mapped to the effect of infinite gravitational red-shift near event horizon and the infinite amount of co-ordinate time the light/particle takes to reach the horizon, starting from a finite distance away from the horizon. Utilizing the dictionary, we elucidate the concept of gravitational red-shift and co-ordinate chart, not covering the whole manifold in a more transparent student-friendly manner without using sophisticated machinery of differential geometry.
\end{abstract}
\maketitle

\section{Introduction and Preliminaries}
Introductory Physics students learn early on that Einstein proposed theories of Special and General Relativity. But the subjects are covered in very different fashion. Even at early stages of College Physics courses the Special theory is presented with care, its foundational principles clearly stated, their mathematical form given, the physical implications elucidated and often the conceptual intricacies explored by considering and resolving various paradoxes. \cite{introductory text 0}  By contrast, it is common to mention the General theory only, at best, in two places:  \cite{inverno}  (i)  in the context of the equivalence principle as it pertains the puzzling equality of the inertial and gravitational masses, appearing in Newton's second law of motion and Newton's theory of gravitation, and (ii) in the context of corpuscular theory of light, the fact that in the General theory there is an associated escape velocity from an astrophysical object for these corpuscules,  is mentioned as a quick peek into the nature of Black holes.Students are left wanting for more specific and better explanations of the content of the General theory, but they are told the mathematics are very involved and best left for a later stage in their education. \\

In this paper we seek for a middle ground.  We attempt to introduce some of the subtleties associated with General Relativity (GR) without necessarily delving into the full machinery of differential geometry. We believe some  fascinating GR stories can be told and explained through carefully chosen analogues from elsewhere in Physics. This can both further whet the early student's appetite for further learning GR and present an opportunity to use various tools from mathematics and physics in a challenging and attractive context.  We present here one such example, an analogy between Zeno's paradox and that of a body falling into a black hole as seen by a falling and a distant observers. 

Zeno's Dichotomy paradox \cite{Field} is presumably well known to students.  To reach his destination, Bob must travel a distance $d$ and he first travels half of it, that is, a distance of  $\frac{d}{2}$.  Then, to travel the remaining distance, $\frac{d}{2}$, he again travels half of it, that is $\frac{d}{4}$. Having traveled a distance of $\frac{3d}{4}$, he next travels half of the remaining distance, or $\frac{d}{8}$. And so on:   at each subsequent ``step'' Bob travels half the remaining distance. Since there is always some distance left to travel,  we must conclude Bob never reaches his destination. This breaking of the trip into artificial steps is arbitrary, but certainly possible and completely general. On the other hand, any beginning physics student knows that  if Bob is traveling with constant velocity $v$, it takes him a time of $\frac{d}{v}$ to reach his destination. This is clearly a finite time, hence Bob indeed reaches his destination. The solution to Zeno's paradox lies in how we reconcile the fact that Bob has to take infinite number of steps to perform the task with the fact Bob takes a finite amount of time to complete it. 

The notion of event horizon of a black hole is probably less familiar. One can introduce it using only familiar concepts from classical mechanics; although not quite correct,  the derivation does give the right answer and more importantly introduces the correct concepts. Students know that the escape velocity $v_{\rm esc}$ from a spherical ``planet'' of mass $M$ and radius $r$ is given by 
\begin{equation}
\label{eq:vescape}
v_{\rm esc}^2=\frac{2G_NM}{r}
\end{equation}
for any body, regardless of its mass. \cite{introductory text} More precisely, this is the escape velocity for a body thrown radially outwards from the ``planet'' from a distance $r$ away form the center. The planet is assumed to be itself spherically symmetric, so all of the mass $M$ could be equivalently considered to be concentrated at the center. This is the situation we wish to consider. Now, the maximum speed an object can attain is that of the speed of light, $c$. Even though this is strictly a limiting velocity ---no massive object can ever attain it--- we can proceed with the argument below by considering the limiting case. For a given ``planet'' (fixed $M$) there is a sphere of radius $R$ from which it takes speed of light to escape; from Eq.~\eqref{eq:vescape} it is given by
\begin{equation}
R=\frac{2G_NM}{c^2}\,.
\end{equation}
Since the speed of light is the maximum speed attainable, any object {\it closer} to the center of the ``planet'' than $R$ will not escape. Even light emitted from the ``planet'' will not escape. This is what is meant by a ``black hole'' and the critical distance $R$ defines a spherical surface called ``event horizon.''

We have surreptitiously invoked the corpuscular theory of light, that posits light as being composed of massless corpuscules, or ``photons.''  They travel at the speed of light and are still subject to the laws of gravitation. The alert student will complain that the photons must slow down as they move away from the planet, but this is impossible since they always move at the speed of light. Here is where GR comes in, in a way markedly distinct from Newtonian gravity. In order to maintain the speed of light fixed, GR declares that the passage of time and the extension of space around the ``planet'' are modified precisely in a way that preserves the speed of light. Instead of the speed of a photon emitted from the event horizon slowing down to zero as the photon escapes to infinity, the wavelength $\lambda$ of the photon is infinitely stretched out as it escapes to infinity, while the frequency $\nu$ is decreased, just in such a way that the speed $c=\lambda\nu$ remains constant. We say that the photons are (infinitely) ``red-shifted'' a term that reflects the fact that red light has the longest wavelength in the visible spectrum.\cite{inverno}

It deserves mention that Zeno's paradox analogy of a Black Hole is not quite perfect. There is no
gravitational field in Zeno's case. Instead the presence of an event
horizon and the redshift of ``escaping'' photons arise because of a
peculiar choice of observers. We are in fact piggybacking on the well
known fact that coordinates that are comoving with accelerated
observers display event horizons and see redshifts. In section II we
give a presentation of the paradox and the gravitational analogy that
may be suitable for representation to students, in section III we
explain, for the benefit of the instructor, the connection in
mathematical detail and in section IV we extend these results to more
general Zeno-motions.

\section{Zeno's paradox revisited}
The solution to Zeno's Dichotomy paradox involves two observations. First, after an infinite number of steps of ever decreasing length, Bob {\it does} make it to its destination. The distance traveled on the $n$-th step is $d/2^n$, so that the total distance travelled is 
\begin{equation}
\sum_{n=1}^{\infty}\frac{d}{2^{n}}=d\,.
\end{equation}
And second observation is to compute correctly the time it takes Bob to reach his destination; while using the step by step approach we need to appreciate that the time elapsed during a step is not universal, but gets smaller and smaller. The time interval over the $n$-th step is $(d/2^n)/v$, so that the total time elapsed is 
\begin{equation}
\sum_{n=1}^{\infty}\frac{1}{v}\frac{d}{2^{n}}=\frac{d}{v}\,.
\end{equation}

Consider now a different situation. ``Bob'' is actually a robot that works, as every computer does, by processing instructions sequentially, with a clock setting the pace of the sequential instructions. So for Bob-the robot, it is simplest to accept that the time interval for any one step is 1 (in arbitrary time units). The instructions in Bob's program still state that he (it?) should move forward by half the distance of the previous step in each consecutive step. Anyone  (technically, ``observer'') that chooses to set her clock to tick according to Bob's steps will record that an infinite amount of time will have to pass in reaching Bob's destination. 

Let's cast this in way that is suitable to build up the Zeno-Gravity correspondence. Let us suppose our robot, Bob,  starts a distance  $r=4$ from a center of a ``planet''  and moves radially with a unit radial velocity ($\frac{dr}{d\tau}=-1$) towards the center at $r=0$. Moreover, suppose Bob's goal is to reach the spherical shell at $r=2$.  This shell can be taken to be a real object, a wall,  {\it e.g.}, the surface of the spherical planet.  If he is able to reach his destination, on reaching $r=2$, he is instructed to continue his journey through a door located on the wall at his destination. Now it can easily be deduced that with unit velocity, Bob hits the wall in $\Delta \tau=2$ units of time. Bob will eventually cross the wall in finite amount of $\tau$-time.\\

As in Zeno's paradox, let's now divide Bob's motion in steps:
\begin{enumerate}
\item In first step he goes from $r=4$ to $r=3$.
\item In $n$-th step, he goes half of what he has traversed in $(n-1)$-th step. 
\end{enumerate}
We have a sequence  of steps given by $\{a_{n}\}=\{1, \frac{1}{2}, \frac{1}{4}, \frac{1}{8}....\}$, where $a_n$ denotes the radial distance traversed in units of $r$ ---or equivalently the time spent in unit of $\tau$--- in the $n$-th step.

Now consider a measure of time agreed upon by those who use the robot's internal clock, one ``step'' is one unit, and denote this by $t$.  That is, $\Delta t=1$ implies one step. While $\tau$ and $t$ are very different measures of time, one is a monotonic function of the other, and in that sense they are both equally good quantities to represent time: both uniquely specify when an event takes place. However, the ``step''  time, $t$, cannot be used to describe any event taking place after Bob crosses the shell at $r=2$. To find the relation  between the two time conventions we can use
\begin{equation}
\label{step}
\tau (t=n)-\tau (t=n-1)=2^{1-n}
\end{equation}
and the condition that both stopwatches are started at the same instant, $\tau(t=0)=0$. Then 
\begin{equation}
\label{step-solved}
\begin{aligned}
\tau (n)-\tau (0)&=\big(\tau (n)-\tau (n-1) \big)+\big (\tau (n-1)-\tau (n-2) \big)\\
&\quad+\cdots+ \big (\tau (1)-\tau (0) \big)\\
&=\sum_{k=1}^n 2^{1-k}\\
&=2(1-2^{-n})\,.
\end{aligned}
\end{equation}
So far the time $t$ has been defined as a discrete quantity. We can interpolate this result for a continuously defined ``step'' time:
\begin{equation}
\label{tau-of-t}
\tau (t)=2(1-2^{-t})\,.
\end{equation}
Notice that this is indeed a monotonic function of $t$. It is however not invertible since there is no  image for $\tau>2$. Similarly one has 
\begin{equation}\label{bm}
r_{n}\equiv r(t=n) = 2\left(1+2^{-n}\right)
\end{equation}
and the $t$-continuum version,
\begin{equation}
\label{bm-continuum}
r(t) = 2\left(1+2^{-t}\right)\,.
\end{equation}

Recall that in this modified version of Zeno's tale, the paradox is that Bob reaches, and then crosses the shell at $r=2$ in finite time $ \tau=2$, but it takes $t\to\infty$ to reach the shell and there is no time $t$ that describes what happens to robot Bob after it reaches the $r=2$ shell. This is the physical content of the result that there is no $t$ for which $\tau(t)>2$. So, does Bob cross the shell at $r=2$?  Now we are ready to use this as analogy of a body (Bob!) falling through an event horizon ($r=2$) of a black hole. The narrative depends very much on the observer. The asymptotic observer ---one very far from the black hole, detecting light emitted form the falling body--- measures time $t$ with his clock ticking uniformly. She is in an inertial frame and Newton's laws of motion hold for her.  And she infers, from looking at Bob falling, that he never reaches the event horizon. Its trajectory is given by Eq.~\eqref{bm-continuum}, which shows that she must wait an infinite amount of time to see Bob reach the horizon. However, an observer close to the horizon, or more precisely, an observer falling into the BH together with Bob, measures with time $\tau$ and concludes that  Bob reaches the horizon in finite time. There is no paradox. Different observers have different narratives, and this is an essential component of the theories of relativity. 

Note that the observer free falling into the BH with Bob concludes Bob's motion is uniform: rectilinear with constant speed. One of the fundamental principles of GR is that inertial frames are defined by free falling observers. Moreover, in GR the ``planet'' changes what we mean by distane and time, but does not exert a gravitational force. Bob's motion described by the observer moving with Bob is uniform because it is described in an inertial frame and there are no forces acting on Bob (hence no acceleration, hence uniform motion). 

Now suppose that Bob is emitting pulses of light at fixed frequency according to the free falling time $\tau$. The rate at which these are observed by the asymptotic observer decreases over time: 
\begin{equation}
\frac{d\tau}{dt}=2\ln(2)\;2^{-t}
\end{equation}
Equivalently, light emitted by Bob is increasingly redshifted as it reaches the asymptotic observer. 

Before closing this section we remark on is one aspect of this analogy which is certain to confuse students. It would have seemed natural to assign time $\tau$ to the asymptotic observer and time $t$ to the free faling one. After all, $t$ is defined as robot-step time so it seems to naturally belong with the robot, while $\tau$ seems to be externally defined so it seems more naturally assigned to the far away observer. However, in GR there is no preferred time coordinate. The closest we can get to a special, preferred time, is time for a free falling observer\cite{footnote}.This is why it is $\tau$ that is the correct time to use for the local description of Bob's free fall in the analogy.

\section{Gravitational System}
The aim of this section is to justify  the analogy for Zeno-Gravity correspondence. This is not material that would be presented to a beginning physics student, but could be used in a higher division introductory GR course. We start by constructing a spherically symmetric metric such that equation of incoming radial null geodesic satisfies \eqref{bm-continuum}. By this we mean, if the (massless) particle starts from $r=4$ at $t=0$, its position at the $n$-th unit of time will be given by \eqref{bm}. It wil be useful to consider the time derivative of \eqref{bm-continuum},
which yields
\begin{equation}\label{master1}
\frac{dr}{dt}=-2\ln(2)\;2^{-t}=-2\left(\frac{r}{2}-1\right)\ln(2)
\end{equation}

Now, to construct the metric we take the following ansatz:
\begin{equation}
\label{metricA}
ds^{2}=f(r)dt^{2}-\frac{1}{f(r)}dr^{2}-k(r,t)d^{2}\Omega_{2}
\end{equation}
where $d^{2}\Omega_{2}$ is the metric of the unit sphere. Although the
setup is spherically symmetric we do not know yet whether the
coordinate $r$ is a the asymptotic radial distance (we will see it is
not); hence the coefficient of the metric for the unit sphere, $k(r,t)$,
is left unspecified for now. From Eq.~\eqref{metricA}, the equation
for the incoming radial null geodesic is given by
\begin{equation}\label{master2}
\frac{dr}{dt}=-f(r)
\end{equation}
Upon comparing \eqref{master1} with \eqref{master2}, we obtain the desired metric:
\begin{equation}
\label{metric}
ds^{2}= 2\ln(2)\left(\frac{r}{2}-1\right)dt^{2}-\frac{1}{2
  \ln(2)\left(\frac{r}{2}-1\right)}dr^{2}-k(r,t) d^{2}\Omega_{2}
\end{equation}

This metric admits an event horizon at $r=2$. Working our way back, the metric gives Eq.~\eqref{master1} for incoming radial null geodesics, with general solution
\begin{equation}\label{rng}
t(r)=-\ln\left(\frac{r}{2}-1\right)/\ln(2)+A\,.
\end{equation}
Imposing the initial condition that $r=4$ at $t=0$, as in robot Bob's motion of the previous section,  sets $A=0$. Inverting this \eqref{rng} gives $r(t)$, indeed reproducing \eqref{bm-continuum}, and of course the step-wise version in Eq.~\eqref{bm}.

The redshift as a photon recedes from the horizon from $r=r_1$ to
$r=r_2$ is computed by standard methods \cite{inverno} and given by 
\begin{equation}
\frac{\omega_2}{\omega_1}=\sqrt{\frac{r_1/2-1}{r_2/2-1}}
\end{equation}
This formula is somewhat confusing because the $\omega_2\to 0$ both
for   $r_1\to 2$  but {\it also} for $r_2\to\infty$. The extreme
redshift $\omega_2\to 0$
observed as the location of photon emission approaches the horizon is
in close analogy with the situation for a photon climbing out of the
gravitational well of a black hole. The redshift as $r_2\to \infty$ may
seem perplexing at first. It stems from the fact that the coordinate
$r$ is not really a good radial coordinate to discuss asymptotic
observers: in fact, for $k(r,t)=r^2$ the metric is
not asymptotically flat. 

Now, the robot's motion is uniform and occurs in flat space. There is
a coordinate system for which this is apparent. To discover this 
we use standard methods: taking a  cue from the equation for a radial,
null geodesic, Eq.~\eqref{rng}, we define a new ``turtoise'' co-ordinate \cite{inverno}
\begin{equation}
r_{*}=\ln\left(\frac{r}{2}-1\right)/\ln(2).
\end{equation}
Then we  define
\begin{equation}
\begin{aligned}
u=t+r_{*}\\
v=t-r_{*}
\end{aligned}
\end{equation}
so that the metric is given in the $(u,v)$ by 
\begin{equation}
ds^{2}= 4\ln(2)e^{\frac{u-v}{2}}dudv-k_{1}(u,v)d^{2}\Omega_2
\end{equation}
where $k_{1}(u,v)=k(r,t)$.
Defining
\begin{equation}
U=2\sqrt{\ln(2)}e^{u},\ \ V=2\sqrt{\ln(2)}e^{-v}
\end{equation}
the metric takes the  form
\begin{equation}
ds^{2}= dUdV-k_{2}(U,V)d^{2}\Omega_2
\end{equation}
where $k_{2}(U,V)=k(r,t)$. Finally, with  $T=U+V$ and $R=U-V$, the
metric takes a form suggesting flat space,
\begin{equation}
\label{RTmetric}
ds^{2}= dT^{2}-dR^{2}-k_{3}(T,R)d^{2}\Omega_2~.
\end{equation}
Here $k_{3}(T,R)=k(r,t)$. Choosing the $k(t,r)$ in a suitable way will
ensure $k_{3}(T,R)=R^{2}$, which will render the metric  flat. The
apparent horizon structure is observed in the original metric form, Eq.~\eqref{metricA}, since it is adapted to an accelerated/decelerated observer.

The in-falling body has an ever decreasing velocity in units of
``step,'' which is the analogue of gravitational red-shift. In context
of gravity, the $(t,r)$ co-ordinate system describes the space-time
for $r\geq2$; it is incomplete, {\it i.e.,} does not cover the whole
space-time. The new coordinates $(u,v)$ cover the same region of
space-time and are unconstrained, that is, $-\infty<u,v<\infty$. This
gives in turn $0<U,V<\infty$ and therefore $0<R\le T<\infty$. But from
Eq.~\eqref{RTmetric} it is clear that nothing peculiar happens at
$R=T$ and in fact one can extend the metric to include arbitrary
time $T$ and arbitrary (positive) radial coordinate $R$. The conclusion
is that the singularity at $r=2$ is a coordinate artifact, a
``coordinate singularity," and the space can be extended past it (yes,
the robot continues to move past $r=2$!).

There is a simpler way to reach the same conclusion. We can define a new time co-ordinate $\theta$ by
\begin{equation}
\label{newtime}
\theta = 2(1-2^{-t})
\end{equation}
in which, the motion of the massless particle is given by $\theta=4-r$. The light reaches the horizon in finite time $\theta$. Moreover, from the explicit form of the metric in $(\theta,r)$ coordinates, 
\begin{equation}
ds^2 = \frac{1}{2 \ln(2)\left(\frac{r}{2}-1\right)}(d\theta^2- dr^{2}-k'd\Omega_{2}^{2})\,,
\end{equation}
where $k'=k/\left[2 \ln(2)\left(\frac{r}{2}-1\right)\right]$, there is no obstacle
to extend the space into $r<2$. The reason is that the singular
factor can be removed by a conformal transformation. In this
coordinate system one can describe the motion of the particle past the
$r=2$ horizon. Hence, $\theta$ is a good co-ordinate to answer the question
of whether the particle reaches the horizon. This $\theta$ co-ordinate
plays a role analogous to that of the time $\tau$ in our robot Bob
example. It should come as no surprise that the motion is
uniform in $(\theta,r)$ coordinates, $dr/d\theta=-1$.

\begin{table}[ht]
\caption{Dictionary}
\centering
\begin{tabular}{|c|c|}
\hline\hline
 Zeno-Robot\ & Gravity\ \\ [0.5ex]
\hline
Bob & light/particle\\
\hline
Bob's motion & Null/Time-like geodesic\\
\hline 
Step & time $t$\\
\hline
$\tau$ & time $\theta$\\ 
\hline
Decreasing velocity in ``Step" unit & Gravitational Redshift\\
\hline
The sum of concerned Series & Position of Horizon\\ [1ex]
\hline
\end{tabular}
\label{dic}
\end{table}

To summarise, we have the dictionary in Tab.~\ref{dic} above.  Zeno's paradox can be thought of arising because we have a bad way of counting time, {\it i.e}, the units of ``step,'' which do not cover the whole spacetime where Bob's motion can take place. By contrast the standard time co-ordinate $\tau$, in which Bob has a constant velocity, is a good way to keep time count. This is analogue to the fact that the co-ordinate time $t$ is not a nice time-coordinate to describe incoming null or time-like geodesic in a
geometry containing an event horizon whereas the newly defined $\theta$ is a good co-ordinate.  \\

\section{Generalisation \& Discussion}
There is nothing special about the sequence $\{1,\frac{1}{2},\frac{1}{4},....\}$ in discussing the Zeno's paradox. Similar situation arises if we have an infinite sequence with a property that the sequence of partial sum of the original sequence converges in $\mathbb{R}$. Hence, there is nothing special about the metric \eqref{metric}. For every such sequence, we can construct a spherically symmetric geometry with an event horizon and the above argument goes through with the dictionary in Tab.~\ref{dic}.  Similarly, for every spherically symmetric geometry with an event horizon, we can construct an infinite sequence and pose the Zeno's paradox.  Here we will work out the general formalism and mention about the underlying assumptions to construct such correspondence.\\

 Let us suppose, we are given with Zeno's paradox, corresponding to a summable sequence $\left\{a\right\}_{n}$ where $\sum a_{n}=M<\infty$ is a finite number. In the robot version, Bob starts from a radial distance of $A>M$, and after nth step, its position is given by, $r_{n}=A-\sum_{k=1}^{n}a_{k}$.  Moroever, we assume that the sequence of partial sum defined by $\left\{S\right\}_{n}$, where $S_{n}=\sum_{k=1}^{n}a_{k}$ can be recast in continuous version as $S(t)=g(t)$, which leads to a velocity $\frac{dr}{dt}=-g^{\prime}(t)$. Assuming $r(t)=A-g(t)$ can be inverted to yield $t$ as a function of $r$, we can express 
\begin{equation}
\label{eq:invert}
g^{\prime}(t)=f(r)
\end{equation}
Given $f(r)$ we can easily construct the metric \eqref{metric},
mentioned in the previous section. Now the presence  of an event horizon
at $r=r_e$ 
is the statement that $f(r_e)=0$. Its existence can be shown in the following way: 
The fact $\sum a_{n}$ converges implies $a_{n}\to 0$ as $n\to\infty$, hence we have $S_{n+1}-S_{n}\to 0 \ \ as\ n\to\infty$, which in turn implies, 
$g^{\prime}(\infty)=0$. Now after an infinite number of  steps, the
robot is at $A-M$. Hence, $r(\infty)=A-M$, and consequently
$f(A-M)=0$: there is an event horizon at $r_e=A-M$.

Conversely, given any geometry and co-ordinate system where we have an
event horizon,  infinite red shifting will occur.  The radial null
geodesic will be described by $r(t)$ that  takes infinite time to
reach the horizon starting from some point outside the horizon. The
radial null geodesic can be sampled uniformly with each internal unit
being  a ``step.'' From there we can easily construct the robot's
motion  and we can construct a summable sequence $\left\{a\right\}_{n}$
such that $A-r(n)= \sum_{k=1}^{n}a_{k}$.\\

One can easily produce from this exercises for the students. If one
aims at closed form solutions, one must take care to find functional
forms of $g(t)$ that makes the inversion, Eq.~\eqref{eq:invert},
posible. A very simple example is $g(t)=\exp(t)$ (and it is very
closely related to Zeno's paradox discussed in the previous
sections). One may also start from the function $f(r)$, and then again a
closed form solution may require careful attention to this choice.
In this class of examples is the actual Black Hole metric. A good exercise for the students is to work out the sequence corresponding to the Schwarzschild metric,\cite{hawking} given by
\begin{equation}
\label{metricS}
ds^{2}=\left(1-\frac{2M}{r}\right)dt^{2}-\left(1-\frac{2M}{r}\right)^{-1}dr^{2}-r^{2}d^2\Omega_{2}
\end{equation}
\\[1pt]
\noindent where $M$ is the mass of the gravitating source as observed by an asymptotic observer, and explicitly show the zeno-gravity correspondence.

\begin{acknowledgments}
The author SP would like to acknowledge a debt of gratitude to Mainak for a discussion on Zeno's Dichotomy paradox on a early winter morning and Prof. Narayan Banerjee for their feedback on the manuscript and Swet for inspiration. This work was supported in part by the US Department of Energy under contract DE-SC0009919.\\
\end{acknowledgments}


\begin{thebibliography}{100}
\bibitem{introductory text 0}Kittel, Knight, Ruderman, Helmholtz, Moyer, Mechanics (Berkeley Physics Course, Vol. 1) 2nd Edition
\bibitem{inverno} D'Inverno. R. \textit{Introducing Einstein's Relativity}, (Clarendon Press, Oxford 1992)
\bibitem{Field}  Field, Paul and Weisstein, Eric W. ``Zeno's Paradoxes."  \htmladdnormallink{From MathWorld--A Wolfram Web Resource.}{http://mathworld.wolfram.com/ZenosParadoxes.html}
\bibitem{introductory text} Giancoli, Douglas C. (2008). Physics for Scientists and Engineers with Modern Physics. Addison-Wesley. p. 201. ISBN 978-0131495081.

\bibitem{hawking} Schutz, Bernard, \textit{A first course in general
  relativity}, Cambridge University Press, 2009.; S. W. Hawking, G. F. R. Ellis, \textit{The large scale structure of space-time}, Cambridge Monographs on Mathematical Physics, Cambridge University Press, p. 149
\bibitem{footnote} Even then it is locally defined, that is, defined only at the location of the observer. Extension beyond the observer is arbitrary.
\end{thebibliography}
\end{document}